# Field-induced Domain Reorientation and Polarization Rotation of <110> Oriented Pb(Mg$_{1/3}$Nb$_{2/3}$)O$_3$-PbTiO$_3$ Single Crystals

Ke-Pi Chen[1,*], Xiao-Wen Zhang[1], Fei Fang[2], and Hao-Su Luo[3]

[1]*State Key Lab of New Ceramics and Fine Processing, Department of Materials Science and Engineering, Tsinghua University, Beijing 100084, P. R. China*

[2]*Department of Engineering Mechanics, Tsinghua University, Beijing 100084, P. R. China*

[3]*State Key Lab of High Performance Ceramics and Superfine Microstructure, Shanghai Institute of Ceramics, Chinese Academy of Sciences, Shanghai 201800, P. R. China*

Abstract：

Polarization hysteresis loops, x-ray diffraction (XRD) and temperature dependent dielectric constant under different electric fields for <110> oriented 0.7 Pb(Mg$_{1/3}$Nb$_{2/3}$)O$_3$-0.3PbTiO$_3$ single crystals were measured. The field-induced phase transition and the process of depolarization were discussed. The results show that with the electric field increasing, the crystal form changes along $R_{relaxor} \to R_{normal} \to M \to O$ via polar-axis reorientation and polarization rotation. However, the depolarization process is not just the reversal of the polarization process. It is noticed that just from the temperature dependent dielectric behavior is not enough to judge the processes of the E-field induced phase transition.

*Email address: chenkepi99@mails.tsinghua.edu.cn

Single crystals of (1-x) Pb(Mg$_{1/3}$Nb$_{2/3}$)O$_3$-xPbTiO$_3$ (PMNT) and (1-x)Pb(Zn$_{1/3}$Nb$_{2/3}$)O$_3$-xPbTiO$_3$ (PZNT) near their morphotropic phase boundary (MPB) are under extensive investigation for their extraordinarily high dielectric and piezoelectric behavior [1-5]. The application of those single crystals may cause breakthrough in ultrasonic transducer materials and devices, as well as the industrial non-destructive inspecting system. Crystallographically, high performances are achieved for <001>-oriented rhombohedral crystals, though <111> is the spontaneous polar direction.

As is well known, piezoelectric behavior of ferroelectric materials reflects the reorientation of ferroelectric domains under external electric field. 180° domain switching will not cause any stain effect, while non-180° domain switching causes sizable strain and consequently influences the apparent piezoelectric constant. In rhombohedral PMNT and PZNT single crystals, there are all together eight <111>-domain states which are randomly distributed. Under E-field, the domains will reorient to align themselves as close as possible to the E-field. Constrained by the surrounding grains, there will be hardly polarization rotation for ferroelectric ceramics, while there is not the case for single crystals. The polarization direction changes under sufficient E-field and causes change in lattice symmetry and structure. This is referred to as E-field induced phase transition. Fu and Cohen [6] reported a first-principle study of the ferroelectric perovskite BaTiO$_3$, demonstrating the different piezoelectric response could be driven by different paths of polarization rotation. Based on this theory, there will be intermediate states of monoclinic (M) or orthorhombic (O) phase occurred during the process of polarization rotation from <111> to <001>. Noheda et al [7-9] revealed the existence of a monoclinic phase in PZT ceramics and a new phase diagram for PZT system was obtained. Guo et al [10] suggested that it is the existence of M phase as a bridging one between R and T (Tetragonal) phases, causes the very high piezoelectric response of PZT. Recent studies are also carried on the composition and phase transition for PMNT and PZNT single crystals near MPB [11-14]. Vanderbite and Cohen [15] expanded the free energy formula to the 8th term, theoretically explained the possibility of the existence of the intermediate M and O Phases (VC model).

For PMNT single crystals near its MPB, comparatively few studies were carried out on the field-induced phase transition and the processes of polarization and depolarization. Viehland et al [16] reported an irreversible transformation between a normal ferroelectric and a relaxor ferroelectric state on <001> oriented 0.7 $Pb(Mg_{1/3}Nb_{2/3})O_3$-$0.3PbTiO_3$ crystals. Lu et al [17] analyzed the phase transition of 0.67PMN-0.33PT single crystals with different poling conditions based on the results of temperature dependent dielectric response. In this letter, we demonstrate that it is not convincible to judge the E-field induced phase transition just from the results of temperature dependent dielectric behavior. In this study, polarization hysteresis loops, x-ray diffraction (XRD) and temperature dependent dielectric constant under different electric fields for <110> oriented 0.7PMN-0.3PT crystals were measured. The field induced phase transition and the process of polarization and depolarization were discussed. The result shows that apart from the polarization rotation (deviation from its spontaneous polarization direction <111>), there is irreversible polar-axis reorientation (switching between the spontaneous polarization directions of crystal <111>) firstly occurred. Both the polar-axis reorientation and polarization rotation play important roles on the materials behavior, but the process of the depolarization is not just a reversal of the polarization.

Single crystals of 0.7PMN-0.3PT were grown by the modified Bridgman method described in ref. [18]. It was reported that the segregation behavior during crystal growth results in compositional inhomogeneity of the PMN-PT single crystals even in the same boule [18]. Therefore in this study only one <110> oriented 5x5x0.5 mm$^3$ specimen was used in all experiments to keep the composition unchanged. Polarization hysteresis loops were measured under different drive electric fields at a frequency of 10Hz, using a standardized ferroelectric test system. The sample was then poled under different E-fields at room temperature 20 minutes for 4 cycles (1st cycle: 2 kV/cm, 2nd cycle: 4 kV/cm, 3rd cycle: 10 kV/cm and 4th cycle: 15 kV/cm). After each poling cycle, the silver electrode was removed and XRD experiments were performed on Rigaka D/max-3B diffractometer using Cu-k$_\beta$ monochromatic radiation ($\lambda$=1.39223Å). Later, the sample was painted with silver electrode again and the dielectric constant was measured as a function of temperature at a heating rate of 4 °C/min using an HP4192A Precision LCR meter at different frequencies (1, 10 and 10kHz). Fig. 1 shows the saturated polarization Ps and remanent polarization Pr changed with E-field for unpoled <110> oriented 0.7PMN-0.3PT crystals. Fig. 2 shows the XRD patterns of {330} diffraction lines for unpoled and poled samples. According to the composition of unpoled 0.70PMN-0.30PT crystals, it belongs to rhombohedral phase. Fig. 3(a) shows the temperature dependent dielectric response with different frequencies of this unpoled sample. There is an apparently frequency dispersion which is consistent with the result reported in [16]. It reveals that there are randomly distributed short-range ordering domains in the unpoled 0.7PMN-0.3PT single crystals. There is an additional evidence to support this point from XRD experiment. For the rhombohedral phase, the {330} reflection is a doublet with the lower-2θ corresponding to the (330) reflection and another corresponding to the (-3 3 0) reflection. As shown in Fig. 2(a), the pattern of {330} diffraction line shows an overlapping profile of two peaks with different 2θ. The two peaks of overlapping profile were separated and each of the two peaks has a broader FWHM (full width of half maximum). As shown in Table I, Δ2θ equals to 0.42°. Taking 0.22° (refers to Table I) as an instrumental breadth, the mean dimension of microdomain about 30 nm was estimated by using the Scherrer equation [19].

The polarization process of the sample under alternative electric field can be deduced from the change of Pr and Ps in Fig.1. Roughly, Fig.1 can be divided into 4 regions. Region I is the preliminary poling stage, both Ps and Pr could hardly measure under AC field less than 2kV/cm. It is suggested that only part of the <111> domains switch 180° with the AC field in this region as sketched in Fig. 4(b). Region II is from 3 kV/cm to 5 kV/cm, Pr and Ps increase rapidly, indicating the reorientation of all the <111> domains to certain [111] direction which are the closest to the external electric field as sketched in Fig. 4(c). In this region, 71° switching causes effective poling effect. From the almost equal values of Ps and Pr, it can also be seen that the switched domains keep their final state irreversibly even E-field equals to zero. Further evidence of this explanation from XRD results will be discussed later. Region III is from 5 kV/cm to 10 kV/cm. With the electric field increasing, Ps and Pr changes nearly linearly and the difference between Ps and Pr increases. From XRD data (shown later) we can infer that the polar-axis changes from <111> to <110> via polarization rotation is underway. However the state is partly reverse while E-field recurs to zero. Region IV is from 10 kV/cm to 15 kV/cm, Ps and Pr keep almost constant. Although it is still in the process of the polarization rotation from <111> to <110>, a comparatively stable intermediate polar state ( M phase) may turn up as sketched in Fig. 4(d). This state is not sensitive to the increasing of AC field from 10 kV/cm to 15 kV/cm at frequency of 10 Hz.

Combining Fig. 2 and Fig. 1, we can see the x-ray diffraction results well support the above deduction of the E-field field induced phase transition. But it should be noted that poling is carried under direct electric field for 20 min, while the polarization hysteresis loops are measured under 10 Hz alternative electric field. The results of corresponding E-field strength under both conditions will not be just the same, but the process of the phase transition should be the same. The X-ray diffraction data for single crystal under different poling conditions are shown in Table I. From Fig. 2(b) and Table 1, it can be seen that for single crystal poled with 2 kV/cm electric filed, there is only one peak for (330) diffraction line but not a doublet as unpoled sample presented. It means not only $180^o$ domain switching occurred but also $71^o$ domain switching occurred. The FWHM of line decreases to $0.26^o$ reveals that the effect of the internal random field which deduces the state of short-range ordering (relaxor ferroelectric rhombohedral phase) [20] can be overcome by the 2kV/cm electric field in substance. The long range ordering domain state (normal ferroelctric rhombohedral phase) can be mainly formed and it is irreversible after the electric filed is removed. For single crystal poled at 4 kV/cm, the $d_{330}$ is 0.09491nm, which equals to the sample poled at 2 kV/cm. It indicates that the crystal state of 4 kV/cm poled sample still keeps rhombohedral form. But the FWHM ($\Delta 2\theta=0.22^o$) of reflection is a little narrower than the 2 kV/cm poled sample, illustrating the polar-axes have perfectly switched to certain direction of {111} which are the closest to the external electric field as sketched in Fig.4(c). Fig. 2(d) and Fig. 2(e) show the X-ray diffraction pattern for single crystal poled at 10 kV/cm and 15 kV/cm. It can be seen that FWHM keeps $0.22^o$, while the peak position changes from $94.35^o$ for crystals poled at 4 kV/cm to $94.30^o$ for crystals poled at 10 kV/cm, and to $94.27^o$ for crystals poled at 15 kV/cm respectively. It is obviously related with the polarization rotation from [111] to [110] as shown in Fig. 4(d) and (e). To avoid the break up of the sample, we didn't increase the strength of poling field further. Although the increasing of the $d_{300}$ is going to slower from 10 kV/cm to 15 kV/cm, it still can't make sure that the sample under 15 kV/cm has been changed to the terminal orthorhombic phase.

The temperature dependent dielectric response, under the condition of zero-field heating after poling in different fields (ZFH aft. FC), will help us a lot to understand the process of depolarization. As shown in Fig. 3(a) for unpoled single crystals, there is only one peak at 132 $^o$C, which is the Curie temperature. Near 124 $^o$C, the frequency behavior of the dielectric constant changes suddenly. There are no frequency dispersion of dielectric constant below 124 $^o$C but do appear again above 124 ℃ for all of the samples after poling in different fields. For simplicity only two patterns of 4 kV/cm and 15 kV/cm samples are shown in Fig. 3(b) and Fig. 3(c). The same phenomenon is also observed in PMN crystals under zero-field heating after field-cooling (ZFH aft. FC) [21]. Thus, this is a phase transition recovering from normal ferroelectric rhombohedral phase to relaxor ferroelectric rhombohedral phase. For samples poled under different electric fields, there are additional peaks around 98 $^o$C and 104 $^o$C. It should be related with the process of depolarization. Although the E-field induced phase transition closely depends on the strength of poling field, the tendency of the temperature dependent dielectric response is almost the same for all of samples poled in different fields. It demonstrates that the depolarization is not just the reversal of the polarization. It is suggested that the peak around 90 $^o$C is related with the depolarization of 180$^o$ domain switching regardless of its initial phase form, because the potential well of 180$^o$ domain switching need to overcome is comparatively small. Therefore it should be appeared firstly. For the peak around 104 $^o$C, it returns to the normal ferroelectric rhombohedral domain state with the randomly distributed <111> polar states. This well explains why the crystals poled at different electric fields have just the same type of temperature dependent dielectric response, and it is also noticed that to judge the process of the E-field induced phase transition just from the temperature dependent dielectric behavior is not enough.

In summary, combining the results of the polarization hysteresis loops, the X-ray diffraction experiments and the temperature dependent dielectric response on heating, the evolution of the E-field induced domain structure and induced phase transition can be clearly understood. With the electric field increasing, the single crystal changes along $R_{relaxor} \to R_{normal} \to M \to O$ via polar-axis reorientation and polarization rotation. However, the depolarization process is not just the reversal of the polarization process. It depends on the potential well, which need to be overcome in the process of the domain structure change and phase transition.

This work was funded by NSFC 59995522.

TABLE I. x-ray diffraction data for <110>oriented 0.7PMN-0.3PT single crystal under different poling conditions.

| Field (kV/cm) | $2\theta$ (°) | $d_{330}$ (nm) | FWHM ($\Delta 2\theta$) |
|---|---|---|---|
| 0 | 94.23 | 0.09500 | 0.42° |
|   | 94.56 | 0.09475 | 0.42° |
| 2 | 94.35 | 0.09491 | 0.26° |
| 4 | 94.35 | 0.09491 | 0.22° |
| 10 | 94.30 | 0.09495 | 0.22° |
| 15 | 94.27 | 0.09497 | 0.22° |

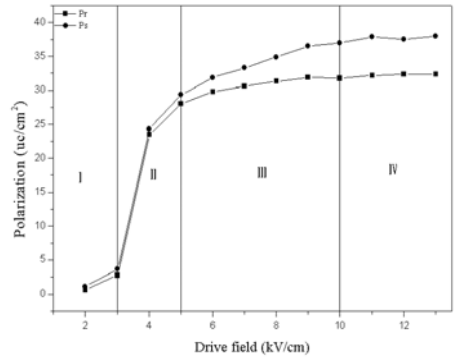

Fig.1 Saturate polarization Ps and remanant polarization Pr change with E-field for unpoled <110> oriented 0.7PMN-0.3PT single crystals.

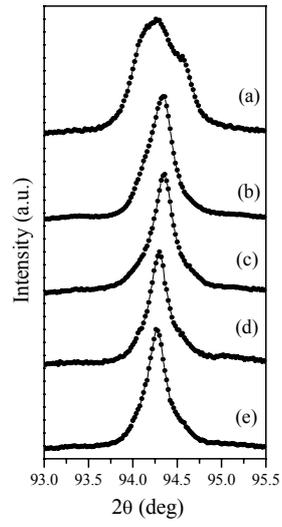

Fig.2 XRD patterns of {330} diffraction lines for 0.7PMN-0.3PT single crystals under different poling fieleds at room temperature for 20minutes (a) unploed (b) $E_{poling}$ = 2 kV/cm (c) $E_{poling}$ = 4 kV/cm (d) $E_{poling}$ = 10 kV/cm (e) $E_{poling}$ = 15 kV/cm

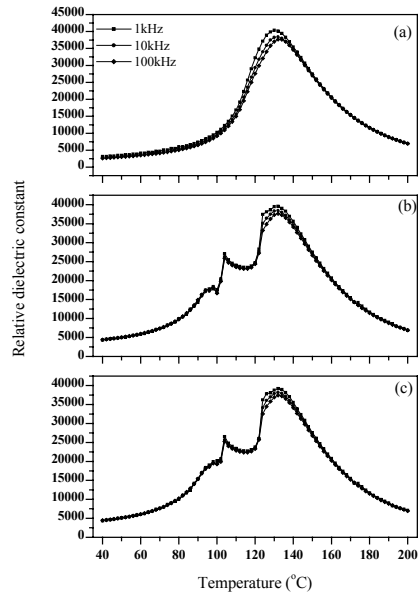

Fig. 3 Temperature dependent dielectric responses for <110> oriented 0.7PMN-0.3PT crystals on heating. (a) unpoled (b) E*poling*=4kV/cm (c) E_poling=15kV/cm

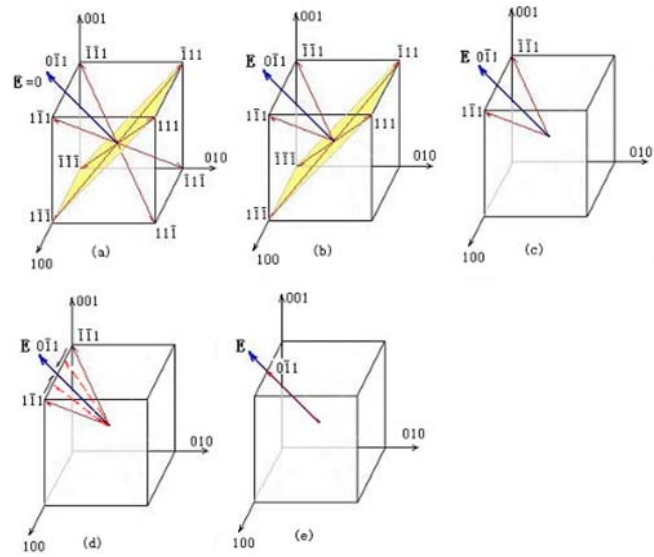

Fig. 4 Schematic view of the change of the polarization axes with electric field increasing.

(a) unpoled state with <111> randomly distributed

(b) only 180° domain switching occurred with no lattice distortion under weak electric field

(c) 71° domain switching occurred with lattice distortion

(d) The two <111> polar axes closest to the E-field rotate toward the E direction with the lattice symmetry change from rhombohedral to monoclinic.

(e) The polarization axis rotate to [110] direction with the lattice symmetry change from monoclinic to orthorhombic.